\documentstyle[manuscript,aps]{revtex}

\newcommand{\beq}{\begin{equation}}
\newcommand{\eeq}{\end{equation}}
\newcommand{\beqa}{\begin{eqnarray}}
\newcommand{\eeqa}{\end{eqnarray}}
\newcommand{\beqar}{\begin{eqnarray*}}
\newcommand{\eeqar}{\end{eqnarray*}}

\def \prob {{\cal P}rob}
\def \la {\langle}
\def \ra {\rangle}

\def \ci {{\cal I}}

\def \ch {{\cal H}}

\def \a {\alpha}

\def \D {\Delta}

\begin{document}

\title{ \bf\Large Trans-Planckian Tail in a Theory with a Cutoff}

\author{B. Reznik \footnote{\it e-mail: reznik@physics.ubc.ca} \\
      {\it Department of Physics \\
      University of British Columbia\\
      Vancouver, B.C., Canada V6T1Z1} }

\maketitle

\begin{abstract}

{
Trans-planckian frequencies can be mimicked outside a black-hole
horizon as a tail of an exponentially large amplitude wave that is mostly
hidden behind the horizon.
The present proposal requires  implementing a
final state condition. This condition involves only
frequencies below the cutoff scale.
It may be interpreted as a condition on the singularity.
Despite the introduction of the cutoff, the Hawking radiation is
restored for static observers.
Freely falling observers see empty space outside the horizon,
but are "heated" as they cross the horizon.

}
\end{abstract}
\newpage

\section{Introduction}

   The standard derivation of the Hawking radiation \cite{hawking}
 requires the existence of exponentially high frequency modes
 in a classical space-time background.
 Indeed after a short time, (of order $t\sim M\ln M$),
 the required frequency becomes  $\omega\sim1$
 in planck units. For this reason
 the  standard derivation can not
 be trusted already  after few Hawking photons were emitted.
It appears that in any derivation of Hawking radiation,
no new physical ingredients,
    a naive short distance cutoff will eliminate the Hawking effect
\cite{jacobson}.
 It is, of course, possible that the origin of the Hawking radiation
 does depend on the behavior of an ultra-high trans-planckian
 spectrum. In this article however we suggest an alternative
  mechanism for generating Hawking's radiation
 in a theory with an effective cutoff.

Evidence that a theory with cutoff may reproduce
the Hawking radiation
has been recently provided by
Unruh's work \cite{dumbhole}.
Unruh has shown
that a natural cutoff still gives rise to the Hawking radiation
in the case of sonic black-holes \cite{sonic}.
  In his approach the cutoff modifies the dispersion
  relation for sound waves.
 This, in turn,
  alters the motion of  modes
  with frequency close to the cutoff scale and gives rise to
  a new type of trajectories which approach the horizon but eventually
  "reflect" back to infinity.
Further works tried to adapt Unruh's model to real black-holes \cite{bpm}
\cite{origin} \cite{corley}.
It is not clear that a similar process is indeed realized for
real black-holes.

   In this article we present another possibility.
It is shown that even without modifying
the ordinary field equations and the ensuing dispersion
relations, as in the above proposals,
one can still restore the Hawking
radiation in a theory with a cutoff.
In the present approach the Hawking radiation is generated by
an apparent trans-planckian tail outside the black-hole horizon.
The source of this tail is an exponentially large wave that is
mostly hidden behind the black-hole horizon.
To develop this picture we shall use
two key ingredients:

\vspace{0.2 cm}

{\it I)  Ultra-high frequency modes can be mimicked to arbitrary accuracy in a bounded region
    even with a finite band spectrum.}

\vspace{0.2 cm}

The basic idea was discovered by Aharonov at. el. \cite{aharonov}
and was further developed by Berry \cite{berry}, who coined
the term "super-oscillations" to describe such a behavior.
A simple  example of a function $F(t)$ which exhibits
super-oscillations
was given in \cite{aharonov}:

\beq
F(t; N, \omega^*) = \biggl[
         \biggl({1-\omega^*/\omega_0\over2}\biggr)e^{it\omega_0/N}
         +\biggl( {1+\omega^*/\omega_0\over2}\biggr)
         e^{-it\omega_0/N}
                \biggr]^N.
\label{func}
\eeq
Here, $N>1$ is an integer, and $\omega^*$ and $\omega_0$ being the
super  and reference frequencies.
For small $t$ we expand $\exp(it\omega_0/N)$ and find:
$$
F(t; N, \omega^*) =
\biggl[ e^{-i\omega^*t/N} + {(\omega^{*2}-\omega_0^2)t^2\over2N^2}
+ O(N^{-3})
\biggr]^N
$$
\beq
=e^{-i\omega^* t} \biggl[   1+   {(\omega^{*2}-\omega_0^2)t^2\over2N} +
O (N^{-2})  \biggr]   \cong  e^{-i\omega^* t}.
\eeq
Although the spectrum of (\ref{func})
includes only modes with frequencies $\omega\in(-1,+1)$,
in the time interval $|t|<< \sqrt{N}/\sqrt{\omega^{*2}-\omega_0^2}\equiv T $,
$F(t)$ behaves as a wave with arbitrary large frequency $\omega^*$. The number
of super-oscillations in this interval is $\sim \sqrt N$.
Systems that interact with the wave
$F$ only during $|t|<<T$
will not distinguish between $F$ and a  pure wave
$e^{-i\omega^* t}$  that  extends for all times.

This remarkable feature is derived at the expense of having such functions grow
exponentially in other regions.
In the example
above, for $|t|>T$, we get $F\sim e^N$.
Nevertheless, as we shall see,
the large amplitudes can be confined to a compact region.
In particular,  by adapting Berry's integral representation\cite{berry},
the large amplitudes can be entirely confined to the interior region
of a black-hole while only a  {\it high frequency tail}
remains outside the black-hole horizon.
This "tail" will be seen by the external observer as
the  origin of the Hawking radiation. The observer can
not probe the interior of the black-hole and can not distinguish
between the mimicked tail and a "truly" trans-planckian
frequency mode.

If the above function F is viewed as a wave function, then
the probability to see a photon coming out from the tail
is exponentially small.
In order to avoid this, we shall make an additional assumption
which is the second basic ingredient:

\vspace{.2 cm}

{\it II) A black-hole is described by  two conditions: by the
ordinary ingoing state and  by a final condition.}

\vspace{.2 cm}

Under this assumption, the black-hole is described in a fashion similar
to that of
a pre and postselected system (\cite{abl},\cite{ts},\cite{ra}).
A pre and postselected ensemble is prepared according to given
initial and final conditions.
Observations can be then made at some intermediate time
between the pre and postselections,
and the probability of the measured results
can be expressed as conditional probabilities.
However in our case the final state will  be given
a more fundamental role.
It will not be determined by a postselection done by some
fictitious observer in the future, rather it
will be conceived as arising from some new fundamental law,
which is required by
 the presence  of a singularity in the future.

Such a  final condition can be anticipated, for
example,  in a  theory that replaces past or future curvature
singularities by smooth initial or final conditions.
To some extent, the Hartle-Hawking ansatz
for the cosmological wave
function \cite{hh} can be interpreted as corresponding
to initial and final
conditions.
When the WKB approximation is valid, the Hartle-Hawking
wave function is expressed in terms of the action $S$ as
\beq
\Psi \simeq e^{iS} + e^{-iS}.
\eeq
It is possible to interpret these two terms  as
two wave functions
which travel forward and backward in time,
and correspond to conditions in  the past or in the future.
It is possible that a similar fundamental principle
is available also for the
case of a black-hole singularity.

Although our assumption  {\it II} above might seem at first radical,
we shall show, in  Section 2,
that final conditions
may be constructed which do not affect low energy observables.
Such a final condition will manifest only in very extreme
cases.
The basic idea will be to implement the condition only on very high
frequency modes above some scale $\omega_h$.

In Section 3 we shall construct the special super-oscillatory
function
which mimics a high frequency tail using a bounded spectrum.
In Section 4 we study the response of a stationary detector
in a  black-hole geometry
to a scalar field when a cutoff with respect to
Kruskal coordinates was introduced.
In section 5 the two main ingredients, namely super-oscillations
and  a final condition, are combined for the
simple case of an eternal black-hole.
A cutoff is assumed with respect to the Kruskal coordinates
both on the initial Kruskal vacuum state  $|O_K\ra$
and our  final state $|f\ra$.
It is shown that these initial and final conditions
cause the observer to see Hawking radiation emitted from the
black-hole.
Finally, we conclude with a discussion of our results and remaining
difficulties.

\section{Final Condition on Ultra-high Modes }

In this section it is shown that a non-trivial final condition
can be imposed without affecting low energy observables.
Over the last decade Prof. Y. Aharonov and collaborators have elaborated
on the two vector formalism of quantum mechanics.
In this formalism one specifies both an initial and a final states
and considers measurements done at intermediate time.
(For a detailed discussion see refs. \cite{ts} and  \cite{ra}.)

Let the initial and final conditions on a system be that at $t=-\infty$ ($+\infty$) the field is in the state $|i\ra$ ($|f\ra$).
Indeed, one can in ordinary quantum mechanics impose two such
conditions. These states are independent,
but need to be non-orthogonal.

In the following we shall consider measurements at some intermediate
time.
Given an observable $A=\sum a\Pi_a$, where $\Pi_a$ are
projectors to the eigenstates $|a\ra$, the probability
to measure $A=a$ is given by the  conditional probability
\footnote{Clearly (\ref{cond}) is different from $|\la a | i\ra|^2$,
the probability obtained if only the initial state is fixed.
The latter is obtained from (\ref{cond}) by further summing over $f$.}:
\beq
\prob(a|f,i) = {\prob(a,f|i)\over \prob(f|i)}=
{|\la f |a\ra\la a |i \ra |^2 \over
\sum_{a'} |  \la f |  a'\ra \la a'| i\ra |^2     }.
\label{cond}
\eeq

For certain non-trivial final conditions,
low energy laboratory experiments will not depend on the final
condition.
In the example considered here
a final condition is imposed  only on
the high energy sector, i.e. only for
$\omega>\omega_h$, where $\omega_h$ is some high
energy scale.

To spell out this proposal, let us for simplicity
consider a free massless scalar field theory
in Minkowski space-time,
and let us assume that in a  certain
rest frame the final state of the field has the form:
\beq
|f_{(L, F)} \ra = {1\over\sqrt{1+\xi^2}} (|L\ra +{\xi} |F\ra),
\label{final}
\eeq
where $|L\ra$ and $|F\ra$ are two normalized states in Fock space,
and $\xi$ controls the relative probability.
The first component, $|L\ra$,
denotes a low energy "laboratory state",
which contains only particles of low  frequency:
\beq
|L\ra = \biggl(1+\sum_{\omega_k<\omega_h}
 C_k(L)a^\dagger_{\omega_k} + \sum_{\omega_k,\omega_l<\omega_h}
D_{kl}(L) a^\dagger_{\omega_k}a^\dagger_{\omega_l} +\dots \biggr)|0_M\ra
\label{L}
\eeq
The second term, $|F\ra$,  denotes a certain state
of particles with frequencies above $\omega_h$:
\beq
|F\ra = \biggl( \sum_{\omega_k>\omega_h} C_k(F)
         a^\dagger_{\omega_k}  + \sum_{\omega_k,\omega_l<\omega_h}
D_{kl}(F) a^\dagger_{\omega_k}a^\dagger_{\omega_l}+ \dots   \biggr)|0_M\ra
\label{F}
\eeq

We shall demand that the final state  always has the
form given in eq. (\ref{final}), and is  $constrained$
always to includes the same high energy state $|F\ra$.
We shall not constrain the content of the low energy
state $|L\ra$.
(In terms of the pre and postselection terminology, this
corresponds to postselection of an ensemble with fixed $F$ but
arbitrary $L$ in the specific combination of eq. (\ref{final}) above.)

Since in the final condition the low energy state $|L\ra$ is
left unspecified, we need to modify eq. (\ref{cond}) accordingly.
The probability to find $A=a$ is  now given by further summing
over a basis of the subspace,
$\ch_L=\lbrace |L\ra\ra\rbrace$,
of low energy states :
\beq
\prob(a| F, i) =  {\sum_{L} P(a,L,F|i) \over
               \sum_{L,a'}P(a', L ,F | i )   }.
\eeq
Thus:
\beq
\prob(a| F, i) =
{\sum_{L}|\la f_{(L,F)} |\Pi_a|i\ra|^2
\over  \sum_{L,a'}
|\la f_{(L, F)} |\Pi_{a'}| i\ra|^2}.
\label{pa}
\eeq

We call $\Pi_a$ a low energy "laboratory" projector if
\beq
\xi ||\la L| \Pi_a|F\ra|| < \epsilon,  \ \ \ \ \ \forall L\in\ch_L
\label{lowe}
\eeq
where $\epsilon$ is some small number. If (\ref{lowe}) is satisfied
for every eigenvalue of the operator $A$, then $A$ will be
termed a  low
energy laboratory observable.
If the initial state is taken to be one of the
low energy states, i.e. $i \in \ch_L$,
then For a low energy observable eq. (\ref{pa}) reduces to
\beq
\prob(a|F, i)= {\sum_{L}|\la L | \Pi_a|i\ra + O(\epsilon)|^2
\over
\sum_{L, a'}  |\la L |\Pi_a'| i\ra +O(\epsilon)|^2   }=
{\la i| \Pi_a|i\ra\over \la i| i\ra} +O(\epsilon),
\label{reduc}
\eeq
the ordinary F independent expression.

Nevertheless, if the initial state does contains
states with $\omega>\omega_h$,
or when the condition (\ref{lowe}) is
not satisfied, the full expression (\ref{pa}) must be used,
and the probability generally depends on $|F\ra$.

Although we have seen that expectation values for low energy laboratory
observables reduce to the ordinary expression,
it is possible
that the fluctuations of the field are still sensitive
to the condition $F$.
To investigate this question let us consider
the case of continuous measurements at intermediate times.
In particular let us
consider the interaction of a particle detector with
the field. This example will be useful in the following sections as well.

A particle detector \cite{unruh76} \cite{dewitt}
can be described as a two level system with an energy gap $\Omega$.
The detector is coupled to a scalar field $\phi(x,t)$ via the action:
\beq
S_I = \lambda \int d\tau dx (A+A^\dagger)\phi(x,t) \delta(x-X_D).
\label{intl}
\eeq
Here, $\tau$ is the proper time in the rest frame of the detector,
$X_D(t)$ is the classical trajectory of the detector.
$A, \ A^\dagger$ act on the two internal states $|\pm\ra$ according to:
$$
A^\dagger|-\ra = |+\ra, \ \ \ \ A|+\ra = |-\ra,
$$
\beq
A^\dagger|+\ra = 0,  \ \ \ \ \ \ A|-\ra = 0.
\eeq
A detection of a particle will be described as a transition
from the ground state to to the  excited state.
In the limit of small coupling constant,
we shall be interested in obtaining the
transition amplitude, computed to the first order in $\lambda$.

With the final condition, the transition probability
is given by:
\beqa
T(+|-,F, i) & = &   { \sum_L
          | \la +, f_{(L, F)}| U_I |-, i \ra|^2
  \over
\sum_{\pm, L }
    |\la  \pm, f_{(L, F)}  | U_I | -, i \ra|^2   }
\eeqa
where $\sum_\pm$ denotes a summation over the final
internal states,
and $U_I=\exp(-i\int L_I d\tau)$
is the unitary evolution operator in the interaction picture.

To lowest order in  the coupling constant
we  get:
$$
   T(+|-,F, i)  =   \lambda^2 {\sum_L
    | \la +, f_{(L, F)}| \int L_I d\tau |-, i \ra|^2
       \over
        \sum_{L}
        \Biggl( | \la\ f_{(L, F)} |i \ra|^2 +
\lambda^2  |\la +, f_{(L, F)}|
             \int L_I  d\tau | -, i \ra |^2 \Biggr) }
$$
\beq
        =  \lambda^2
   \sum_L | \la +| \otimes ( \la L| + \xi \la F|)
   \int L_I d\tau | -, i\ra \ |^2
   + O(\lambda^3) ,
\label{transit}
\eeq
where in passing to the last line we used the identity
$\la F|i\ra = 0$.

The transition probability obtained in (\ref{transit})
has the ordinary form, except that now it
contains the additional  component
\beq
A_F =\xi \int d\tau \la +, F |L_I |-,i\ra.
\label{af}
\eeq
 When $A_F$
vanishes eq. (\ref{transit}) reduces to the ordinary
transition probability.

Let us now consider the  new amplitude $A_F$.
Using the representation (\ref{F}) for $|F\ra$, and neglecting
possible
multi-particle contributions we obtain
\beq
A_F = \xi \sum_k { C^*_k(F) \over\sqrt{4\pi\omega_k}}
\int_{-\tau_0}^{\tau_0} d\tau
                \exp(i\Omega\tau)\exp(i\omega_k(t(\tau)-kX_D(\tau))=
\sum_{\omega_k}A_{F}(k).
\eeq
For an inertial detector, $t = \tau/\sqrt{1-V_D^2}$,
and we find that
\beq
A_{F}(k) = \xi {C_k^*(F)\over\sqrt{4\pi\omega_k}} {\sin\biggl[
(\Omega+ (\omega_k-C_Dk)/\sqrt {1-V_D^2})\tau_0\biggr]
\over
(\Omega+ (\omega_n-V_Dk)/\sqrt {1-V_D^2})\tau_0 }
\eeq
The last eq. reduces to
$\delta(\Omega+ (\omega_n-V_Dk)/\sqrt {1-V_D^2}) =0$
only when  $\tau_0 >> 1/\omega_k> 1/\omega_h$, and provided that
$\xi$ is finite, say $\xi \sim 1$.
This means that as long as the relative amplitude $\xi$
of $|F\ra$ is not large, the fluctuations are averaged out
to zero after a time which is determined by $1/\omega_h$.
Intuitively this seems natural.
An interaction on a time scale shorter  then $1/\omega_h$
involves energy fluctuations of order $\sim \omega_h$,
which in turn depend on the condition $F$.

Anticipating the discussion in Section 5, let us also consider
the large $\xi$ case.
By insisting that eq. (\ref{lowe}) is satisfied
we find that the expectation value (\ref{reduc}) are still
unmodified.
Nevertheless, the fluctuating $A_F$
seen by a particle detector are not negligible.
In this case, to average out such fluctuations
we will need  times $\tau_0 >> \xi/\omega_h$.
Otherwise our detector will observe particles which are not present
in the initial state $|i\ra$.

Finally we note that the above considerations can be easily
extended to the case of a final mixed state.
The analog of the state $|f_{(L, F)}\ra$
is given by the density matrix
\beq
\rho_f = |L\ra\la L| + \rho_{F},
\eeq
where  $\rho_{F_c}$ is constructed from
states with frequency $\omega>\omega_h$.

\section{Ultra-high Frequency from a Bounded Spectrum}

The other key element {\it I} of our approach is the use
of super-oscillatory functions, alluded to in the introduction.
These functions,
having only a bounded Fourier spectrum,  can still mimic an arbitrarily high frequency though in a finite region.

In the Fourier representation of such a function
\beq
\Phi(u) = \int_0^1 d\omega  \ C_\omega \exp{i\omega u},
\eeq
the trick is to choose certain coefficients
$C_{\omega}$, such that at a finite interval  of $u$,
$\Phi$ exhibits  rapid oscillations with a  frequency
$\omega^* >> 1$.

As super oscillations necessitate
large amplitudes at other regions, our purpose is to find
a representation in which these large oscillations
 are confined to a bounded region
of $u$.
To construct such a function we will use a variant of
an integral representation for a  super-oscillatory function
that was found by Berry \cite{berry}.
Consider the function:

\beq
\Phi_{A,\D}(u) = {1\over\sqrt{2\pi \D^2}} \int_0^{2\pi}d\alpha
                e^ { {i\over\D^2} \cos(\a-iA) }
                e^{i\cos\a u}
\label{phi}
\eeq
where $A$ and $\Delta$ are real parameters.
The modes of $\Phi_{A,\Delta}(u)$ are bounded by $|\omega|= |\cos\alpha|<1$.

The  integral  above can be analytically performed to yield
\beq
\Phi_{A,\D}(u)=
    \sqrt{2\pi/ \D^2}  \ I_0 \biggl(
    {1\over\D^2}\sqrt{1+2\cosh(A)\D^2 u +\D^4 u^2}
                    \biggr),
\label{bessel}
\eeq
where $I_0$ is the zeroth modified Bessel function.

Expanding (\ref{bessel}) around $u=0$,
we note that $\Phi_{A, \D} (u)$ behaves as
\beq
      \Phi_{A,\D} (u)= \exp( i\cosh(A) \  u ).
\label{super}
\eeq
$\Phi_{A,\D}(u)$ "super-oscillates"  with
frequency $\omega^*=\cosh A$.
This expansion is valid
in a region  $|u| < (\cosh(A)\D^2)^{-1}
\equiv \delta u$.
Thus the  parameter $\D$  controls the number, $n_S$,
of super-oscillations: $n_S \sim \delta u/\omega^* = 1/\D^2$.

By modifying the two parameters $A$ and $\D$, we can control
the frequency and number of super oscillations.
However, the limits $A\to \infty$ or $\D\to 0$ are singular.
Outside the region $\delta u$ where the function super-oscillates,
$\Phi$ grows exponentially.
$\Phi$ gets its maximal value at
$u=-\cosh A/\D^2$, where the amplitude grows
to
\beq
\Phi \sim \exp({\cosh A/\D^2}).
\label{expamp}
\eeq
The super-oscillations are hence found at the tails  of an
exponentially high peak.

Far away from the region of super-oscillations,
for $u>>\delta u$, (\ref{bessel}) reduces to a low frequency wave:
\beq
\Phi_{A,\D}(u) = {1\over u} \exp(iu)
\eeq

In Section 5 we will show that these properties
of $\Phi$, allow finding a state $|F\ra$
that  mimics the trans-planckian Hawking photons close
to the horizon.
A particle detector will not distinguish between a "fake"
tail of super-oscillations and "real" trans-planckian model.
Before proceeding to this final task, it will be useful to
re-examine the interaction of a particle detector
with a scalar field near a black-hole.

\section{ Particle Detector in Kruskal geometry
with a Cutoff}

In this section we  examine the response of a particle detector
in the space-time of an  eternal black-hole.
We shall assume that
the initial state is the Unruh vacuum \cite{unruh76}, but that modes above
a certain frequency are cutoff. In this section we use usual, only preselected,
quantum mechanics framework. (We shall defer the discussion
of an additional final condition to the next section.)

The geometry of an eternal black-hole is
described in terms of Kruskal
coordinates $U, \ V$,
that are defined via the relations:

$$
ds^2 = ( 1- 2M/r )dt^2 - ( 1- 2M/r)^{-1}dr^2
-r^2 d\Omega^2
$$
$$
=  ( 1- 2M/r )dudv -r^2 d\Omega^2
$$
\beq
= {\exp(-r/4M)\over r/2M} dUdV - r^2d\Omega^2
\label{kruskal}
\eeq
Here,
\beq
u,v = t\pm r^*,
\label{uv}
\eeq
where $r^*= r + 2M\ln(r/2M -1)$, and
\beq
U = -4M\exp(-u/4M),  \ \ \ \ V=4M\exp(v/4M).
\label{UV}
\eeq

The Unruh vacuum corresponds to an initial state  which reproduces
Hawking radiation at $\ci^+$.
It is imposed by selecting the in-vacuum with
respect to the Killing vector $\partial_U$ on the past horizon
($V=0$) as follows.
Consider a scalar field in the reduced 1+1 spherical approximation.
By the conformal invariance we have
\beq
\phi(U,V) = \phi_R(U) + \phi_L(V).
\eeq
In terms of creation and annihilation operators we have
\beq
\phi_R(U) = \int {d\omega\over\sqrt{4\pi\omega}} \biggl(
e^{-i\omega U} a_\omega^R + h.c. \biggr),
\label{rphi}
\eeq
and a similar expression of the left moving part $\phi_L$.

Now for the Unruh vacuum $|0_U\ra$
\beq
a_\omega^R|0_U\ra = 0.
\eeq
Here we furthermore assume that modes with  $\omega>\omega_c$ are cutoff.

Let us consider now a static particle detector
that is located at a constant radius $r$ and interacts with the
cutoff vacuum state defined above.
The trajectory $(r,t)$ of the detector,
can be described in terms of  the Kruskal coordinates
$(U,V)$ as
\beq
U_D= -(r/2M)^{1/2}l\exp[(-t+r)/4M],
\ \  V_D= (r/2M)^{1/2}l\exp[(-t-r)/4M],
\label{uvd}
\eeq
where  $l\equiv 4M(1-2M/r)^{1/2}$.

Using the interaction (\ref{intl}), and eqs. (\ref{rphi},\ref{uvd}),
we shall obtain the transition amplitude from an initial vacuum
state an unexcited detector to a final state with excited detector state
and a one scalar photon of frequency $\omega$:
\beq
A(+, \omega|-, 0_K) = \lambda\int d\tau
\la +, 1_\omega| (A+A^\dagger)\phi(U_D) |-, 0_U\ra,
\label{ak}
\eeq
where the time coordinate is related to the proper time by
$dt= d\tau/(1-2M/r)^{1/2}$.
We find that:
\beq
A(+, \omega|-, 0_U) = \lambda^2
\int  {d\tau \over\sqrt{4\pi \omega_k}}
e^{i\Omega\tau}\exp\biggl(
i\omega_k (r/2M)^{1/2}l\exp[(-t(\tau)+r)/4M]   \biggr),
\label{akrus}
\eeq

The total probability of jumping  to an exited state is obtained
by summation over the final emitted photons states:
\beq
\prob(+) = \sum_{\omega<\omega_c} | A(+, \omega) |^2
\label{tprob}
\eeq

By inspecting the transition
amplitude eq. (\ref{akrus}), one finds that the
integral is dominated by a stationary point at
\beq
\omega_{s.p} =
-4M\exp[(t(\tau)-r)/4M]\biggl( {2M\over r} \biggr)^{1/2}
\eeq
Since the maximal frequency is $\omega_c$, after u-time of order
$t-r \sim 4M\ln(\omega_C/4M)$ this transition amplitude vanishes.
In other words, the emission seen by the detector will
come to halt very shortly after it started.
This corresponds to the usual result that a cutoff
will terminate the Hawking radiation.

\section{Restoring the Hawking radiation}

We shall now show that a particular choice for
the final condition,  gives rise to an effective
ultra-high frequency in the vicinity of the horizon
and avoids the above "extinction" of the Hawking radiation due to the
cutoff.

As we have seen in Section 2, when a final condition on the
high frequency modes is imposed,
the transition amplitude (\ref{transit})  contains an
extra term
(\ref{af}).
The contribution of this term
to the transition amplitude is
\beq
A_F = \lambda \int {d\tau}
\la +, F| L_I |-, 0_U\ra
\label{afk}
\eeq

In order to replace the contribution of a trans-planckian frequency $\omega^*$
by a super-oscillation, we shall require that
\beq
\la +, F_{\omega^*,\D}| L_I| -, 0_U\ra =
    \la +, \omega=\omega^* | L_I | -, 0_U \ra
\label{cons}
\eeq

We shall  assume that the  final state contains only
modes with  frequencies $\omega\in (\omega_c,
\omega_c-\zeta)$, with $\omega$ in $M_{PL}$ units.
$\zeta<1$ is some pure number that defines
the size of the  high energy "band" $below$ the cutoff scale.
Using eqs. (\ref{F},\ref{phi}), together with the
constraint (\ref{cons}),
we can find the coefficients  $C_k(F)$
of the single particle states  in (\ref{F}).
(In this article we shall not construct the
coefficients of  multi-particle terms.)
The result is:
\beq
|F_{\omega^*}\ra =
\int_0^{2\pi} d\alpha \sqrt{\omega_\alpha\over\omega^* }
           \biggl[    \exp{{i\over\D} \cos(\alpha-iA)} \biggr]
              a^\dagger_{\omega_\alpha} |0_U\ra,
\label{fstate}
\eeq
where
\beq
\omega_\alpha = \omega_c + \zeta(\cos\alpha+1)/2,
\ \ \ \omega^*=\cosh A.
\eeq

Outside the black-hole horizon, the effective transition amplitude
(\ref{afk}), (with a sufficiently large number
of super-oscillations $n_S\sim1/\D$),
is precisely identical to that obtained without the cutoff.
Therefore, we have shown that by a superposition of final states
with $\omega \in \omega_c\pm1$, we can mimic a
 trans-planckian  frequency $\omega^*=\cosh A >> \omega_c$.

Freely falling detectors are equivalent to  inertial detectors
in Minkowski space-time. We have seen in Section 2,
that inertial detectors  with a small boost factor,
will not respond to the final condition.
In close analogy, freely falling detectors outside the black-hole
respond only to low frequencies modes.
Thus, a freely falling detector outside the black-hole
effectively interact only with a normal wave
and hence sees the space as mostly empty;
precisely as in the ordinary picture.
Nevertheless, the standard picture  fails at
the interior of the black-hole.
Inside the black-hole
the effective trans-planckian tail
rises sharply to a tremendous amplitude  (\ref{expamp}) of $\sim\exp(\omega^*/\D^2)$.

Although the expectation values of low-energy observable
may remain unchanged, the {\it fluctuations} become
exponentially large.
This implies that a probe that couples during a finite time
to the field will detect particles with high probability.
In the standard picture the black-hole
is basically empty, and nothing extrordinary occurs when
an observer crosses the horizon.
In our case, we expect that as the probe crosses the horizon
it will be immediately start heating up!
We see that in this scenario,
while outside the black-hole the ordinary predictions are
respected,  new physics is predicted
for the region hidden by the horizon.

So far we have only demonstrated that a  single trans-planckian
mode can be restored.
To restore the full transition probability  (\ref{tprob})
we need to mimic the full trans-planckian spectrum.
Since the different frequencies superpose with no interference,
we need to chose a final state which is a density matrix:
\beq
\rho_f = |L\ra\la L| + \int_0^{\omega_{max}}d\omega^*
\ |F_{\omega^*}\ra\la F_{\omega^*} |.
\label{fden}
\eeq
The first term represents the non conditioned final low energy
state, which allows to restores ordinary low energy physics
outside the black hole.
The second term represents the final condition on high energy
states near the cutoff $\omega_c$ and is responsible for the
Hawking radiation.
Strictly speaking, the state $|F_{\omega^*}\ra$
become ill defined in the
limit $\omega^*\to\infty$.
Yet during every finite life time we have a well defined expression.
For example, for an  isolated black hole
that evaporates according the standard picture during $t\sim M^3$,
we will require  in (\ref{fden})
$\omega_{max}\sim\exp M^2$.

\section{Discussion}

In this article we have presented a novel picture in which
the Hawking radiation can be restored without trans-planckian modes.
The source of the radiation in this picture is a
tail of trans-planckian oscillations near the horizon,
that reaches out from a wave with exponentially high amplitudes
hidden inside the black-hole.
Consequently, the interior
of the black-hole becomes "hot" for a freely falling
observer that crosses the horizon.
We therefore expect large back-reaction effects that drastically
modify the interior geometry.

Another new aspect of our approach is the final condition
on the high energy sector.
An appealing motivation for this new
physical condition is the existence of a future singularity.
The condition may be a result of a new physical principle
that replaces the future singularity
by a final state $|F\ra$.

The final state of the high energy sector near the cutoff energy
seems to be  a density matrix.
A possible explanation for this could be the following.
In our approach gravity is essentially treated semi-classically.
If indeed the final state is related to the singularity,
we can not expected it to be a expressed as direct product
of a matter states and a semi-classical gravity state.
Rather it should be a highly entangled
matter-gravity state. Since outside the black-hole
the semi-classical approximation is valid,
the matter state should result from tracing over gravity states.
This procedure may lead to a reduced density matrix like in eq. (\ref{fden}).

The mechanism presented in this paper is far from been
complete, and many important questions remain.
For example,  can F states also mimic the two-point correlation
functions?
Or how does the radiation energy transfer in this picture?
We hope that some of the features presented in this approach
will turn out useful
in understanding of the enigma of Hawking radiation.

\newpage

\vspace {2 cm}

{\bf Acknoledgment}
I am grateful to W. G. Unruh for many helpful discussions
during the preparation of this work.
I also like to thank Y. Aharonov, S. Massar and S. Nussinov.
for helpful discussions and remarks.

\vspace {2 cm}

{\bf Note Added}

After the completion of this work I found out that a
concept related to ingredient {\it I}
was suggested also by H. Rosu \cite{rosu}.

\end{document}